\documentclass[draft,eqsecnum,nofootinbib,aps]{revtex4}
\renewcommand{\theequation}{\arabic{equation}}
\def\beq{\begin{equation}}
\def\eeq{\end{equation}}
\def\bea{\begin{eqnarray}}
\def\eea{\end{eqnarray}}

\begin{document}
\title{Proton strange form factors and SAMPLE experiments}
\author{Soon-Tae Hong}
\email{soonhong@ewha.ac.kr}
\affiliation{Department of Science
Education, Ewha Womans University, Seoul 120-750 Korea}
\date{July 18, 2003}
\begin{abstract}
We study the chiral models such as the Skyrmion and chiral bag to
investigate the baryon strange form factors associated with the SAMPLE 
experiments.
\end{abstract}
\maketitle

\section{Introduction}
\setcounter{equation}{0}
\renewcommand{\theequation}{\arabic{section}.\arabic{equation}}

Recently, the SAMPLE Collaboration reported the
experimental data of the proton strange magnetic form factor through parity violating
electron scattering at a small momentum transfer $Q_{S}^2 = 0.1~{\rm 
(GeV/c)}^2$~\cite{sample01} 
\beq
G_{M}^{s} (Q_{S}^2)=+0.14 \pm 0.29~{\rm (stat)} 
\pm 0.31~{\rm (sys)}~{\rm n.m.}.
\label{data}
\eeq
On the other hand, baryons were described by topological 
solitons~\cite{skyrme61,witten83npb,brown83prl,hong98,hong02pr} and the MIT 
bag model~\cite{chodos74} was later unified with the Skyrmion model to yield the chiral 
bag model (CBM)~\cite{gerry791}, which then includes the pion cloud degrees of freedom and the 
chiral invariance consistently.  Moreover, the soliton  was exploited to yield 
superqualiton~\cite{hong01} in color flavor locking phase~\cite{rajagopal00}.

The QCD is the basic underlying theory of strong interaction, from which low energy hadron 
physics should be attainable.  Moreover, for hadron structure calculations, the 
coupling constant $g$ is not a relevant expansion parameter of QCD.  Long ago, 't Hooft noted 
that $1/N_{c}$ could be regarded as expansion parameter of QCD~\cite{thooft74} where $N_{c}$ 
is the number of colors and $gN_{c}^{2}$ is kept constant.  The properties of large $N_{c}$ limit 
of the QCD can be satisfied by the meson sector of the nonlinear sigma model such as the 
Skyrmion model.  

Now, in order to study the hadron physics phenomenology, we start with the SU(3) 
Skyrmion lagrangian of the form~\cite{witten83npb}
\beq
L=\int{\rm d}^{3}x~\left[-\frac{f_{\pi}^{2}}{4}
{\rm tr}(l_{\mu}l^{\mu})+\frac{1}{32e^{2}}{\rm tr}[l_{\mu},l_{\nu}]^{2}\right]+L_{WZW}
\eeq
where $l_{\mu}=U^{\dagger}\partial_{\mu}U$ and $U\in$ SU(3) is described by 
pseudoscalar meson fields $\pi_{a}$ $(a=1,2,...,8)$ and the topological 
aspects can be included via the WZW action~\cite{witten83npb}.  Assuming maximal 
symmetry, we introduce the hedgehog ansatz $U_{0}$ embedded in the SU(2) 
isospin subgroup of SU(3) to yield the topological charge 
\beq
Q=-\frac{1}{2\pi}\chi_{E}(\theta-\sin\theta\cos\theta)=1
\label{charge}
\eeq
where $\theta$ is the chiral angle and $\chi_{E}$ is the Euler characteristic being 
an inter two in the spherical bag surface.  

\section{BRST symmetries in Skyrmion}

In order to define the spin and isospin one can quantize, in the SU(2) Skyrmion for instance, 
the zero modes via $U_{0}\rightarrow AU_{0}A^{\dagger}$ and $A(t)=a^{0}
+i\vec{a}\cdot\vec{\tau}$ with $a^{\mu}$ being the collective coordinates.  One can then 
obtain the Lagrangian $L=-m_{0}+2i_{1}\dot{a}^{\mu}\dot{a}^{\mu}$ 
where the static mass $m_{0}$ and the moment of inertia $i_{1}$ are calculable 
in the Skyrmion model.  Introducing the canonical momenta $\pi^{\mu}$ one can obtain 
the canonical Hamiltonian $H=m_{0}+\frac{1}{8i_{1}}\pi^{\mu}\pi^{\mu}$.  Note that 
the second-class geometrical constraints 
\beq
\Omega_{1}=a^{\mu}a^{\mu}-1\approx 0,~~~\Omega_{2} = a^{\mu}\pi^{\mu}\approx 0
\eeq
should be treated via the Dirac 
brackets~\cite{dirac64}.  However, in the Dirac quantization, one has difficulties in 
finding the canonically conjugate pair, which were later overcome~\cite{bft} by 
introducing pair of auxiliary Stuckelberg fields $\theta$ and $\pi_{\theta}$ with 
$\{\theta, \pi_{\theta}\}=1$.  In the Skyrmion the first-class constraints $\tilde{\Omega}_{1}
=a^{\mu}a^{\mu}-1+2\theta$ and $\tilde{\Omega}_{2}=a^{\mu}\pi^{\mu}-a^{\mu}
a^{\mu}\pi_{\theta}$ were constructed~\cite{hong99prd} to satisfy the strongly involutive 
Lie algebra $\{\tilde{\Omega}_{1},\tilde{\Omega}_{2}\}=0$.  Similarly, the first-class Hamiltonian 
was formulated to yield the baryon mass spectrum 
\beq
m_{B}=m_{0}+\frac{1}{2i_{1}}\left[J(J+1)+\frac{1}{4}\right]
\eeq
with the isospin quantum number $J$.
Here note that an additional global shift is due to the Weyl ordering correction.  Following the 
BRST quantization scheme~\cite{brst} with (anti)ghost and their Lagrangian multiplier fields, we 
obtain the BRST symmetric Lagrangian~\cite{hong99prd},
\beq
L_{eff}=-m_{0}+\frac{2i_{1}\dot{a}^{\mu}\dot{a}^{\mu}-\dot{\theta}\dot{b}}{1-2\theta} 
-\frac{2i_{1}\dot{\theta}^{2} }{(1-2\theta)^{2}}
-2i_{1}(1-2\theta)^{2}(b+2\bar{c}c)^{2}+\dot{\bar{c}}\dot{c}
\eeq
invariant under the transformations, $\delta_{\lambda}a^{\mu}=\lambda a^{\mu}c$, 
$\delta_{\lambda}\theta=-\lambda(1-2\theta)c$, $\delta_{\lambda}\bar{c}
=-\lambda b$ and $\delta_{\lambda}c=\delta_{\lambda}b=0$.  (For more details of 
the BRST quantization of the SU(2) and SU(3) Skyrmions, see~\cite{hong99prd} and 
\cite{hong01prd}, respectively.)

\section{Baryon strange form factors}

Next, we consider the CBM which is a hybrid of two different models: 
the MIT bag model at infinite bag radius on one hand and Skyrmion model at vanishing 
radius on the other hand.  (The explicit CBM lagrangian is given in
Ref.~\cite{hong02pr} for instance.)  In the CBM the total topological
charge $Q$ in Eq. (\ref{charge}) is now splitted into the meson and
quark pieces to satisfy the Cheshire cat principle~\cite{ccp}.
Moreover, the quark fractional charge is given by sum of integer one 
(from valence quarks) and the quark vacuum contribution, which is also 
rewritten in terms of the eta invariant~\cite{atiyah}.  

In the collective quantization of the CBM, we explicitly obtain the
proton magnetic moment~\cite{hong93,hong93npa}
\beq
\mu_{p}=\frac{1}{90}(9I_{1}+24I_{2}+12I_{3}+16I_{4}-4I_{5}) 
+\frac{2I_{6}}{1125}\left(9I_{1}+4I_{2}-8I_{3}\right) 
\eeq
with the inertia parameters $I_{n}$ $(n=1,...,6)$ calculable in the CBM.  Similarly we 
construct the baryon octet magnetic moments to reproduce 
the Coleman-Glashow sum rules~\cite{coleman,hong93npa} such as $U$-spin symmetries, 
$\mu_{\Sigma^{+}}=\mu_{p}$, $\mu_{\Xi^{0}}=\mu_{n}$ and $\mu_{\Xi^-}=\mu_{\Sigma^{-}}$.

Now we define the Dirac and Pauli EM form factors via 
\beq
\langle p+q|\hat{V}^{\mu} |p\rangle = \bar{u}(p+q)\left[F_{1B}(q^2)
\gamma^{\mu}+\frac{i}{2m_B}F_{2B}(q^2)\sigma^{\mu\nu} 
q_\nu\right]u(p)
\eeq
where $q$ is momentum transfer and 
$\sigma^{\mu\nu}=\frac{i}{2}(\gamma^{\mu}\gamma^{\nu}-\gamma^{\nu}\gamma^{\mu})$ and 
$m_{B}$ is baryon mass.  The Sachs form factors are then given by $G_{M}=F_{1B}+F_{2B}$ and 
$G_{E}=F_{1B}+\frac{q^{2}}{4m_{B}^{2}}F_{2B}$ so that, at zero momentum transfer, 
the Pauli strange form factor is identical to the Sachs strange form factor: 
$F_{2B}^{s}(0) =G_{M}^{s}(0)$.  In the SAMPLE experiment, they measured the neutral 
weak form factor 
\beq 
G_{M}^{Z,p} = \left(\frac{1}{4}-\sin^{2} \theta_{W}\right)G_{M}^{p}-\frac{1}{4}G_{M}^{n}
-\frac{1}{4}G_{M}^{s}
\eeq
with $G_{M}^p$ and $G_{M}^n$ being the proton and neutron Sachs form factors, to predict 
the proton strange form factor (\ref{data}) which is positive value contrary to the 
negative values from most of the model calculations except the predictions~\cite{hong93,hong97plb} 
of the SU(3) CBM and the recent predictions of the chiral quark soliton model~\cite{kim98} 
and the chiral perturbation theory~\cite{meissner00,vankolck}.  (See~\cite{hong02pr} for more details.) 
In the CBM the proton strange form factor is given by~\cite{hong93}
\begin{table}[t]
\caption{The baryon octet strange form factors in the CBM}
\begin{center}
\begin{tabular}{lrrrr}
\hline
Input &$F_{2N}^{s}(0)$ &$F_{2\Lambda}^{s}(0)$ &$F_{2\Xi}^{s}(0)$ 
      &$F_{2\Sigma}^{s}(0)$\\ 
\hline
CBM  &0.30   &0.49 &0.25 &$-1.54$\\ 
Exp  &0.32   &1.42 &1.10 &$-1.10$\\ 
\hline\\
\end{tabular}
\end{center}
\end{table}
\beq
F_{2N}^{s}(0)=\frac{1}{60}(21I_{1}-4I_{2}-2I_{3}-4I_{4}-2I_{5})
+\frac{I_{6}}{2250}(-129I_{1}+76I_{2}-52I_{3})
\label{f2n0}
\eeq
which, after some algebra with the other baryon octet strange form factors, yields 
the sum rule for the proton strange form factor in terms of the baryon octet magnetic 
moments only (for the other baryon sum rules see~\cite{hong01hep}) 
\beq
F_{2N}^{s}(0)=\mu_{p}-\mu_{\Xi^{-}}-(\mu_{p}+\mu_{n})
-\frac{1}{3}(\mu_{\Sigma^{+}}-\mu_{\Xi^{0}})
+\frac{4}{3}(\mu_{n}-\mu_{\Sigma^{-}}).
\label{f2n1}
\eeq
Explicitly calculating the inertia parameters $I_{n}$ numerically in (\ref{f2n0}), 
we predict the proton strange form factor, 0.30 n.m. and exploiting the experimental 
data for the baryon octet magnetic moments in (\ref{f2n1}) we obtain 0.32 n.m..  
These predictions are comparable to the SAMPLE experimental data (\ref{data}) and 
are shown in Table 1, together with those of the other baryon strange form factors.

\section{Conclusions}

In conclusion, we discussed the SAMPLE experiments in the topological solitons such as 
the Skymion and chiral models to predict baryon strange form factors.  We also exploited 
the Dirac quantization associated with ghost field degrees of freedom to construct 
the BRST invariant chiral Lagrangian.   In future, it will be interesting to consider 
in these chiral models anapole form factors discussed in the literatures~\cite{vankolck}.\\

\vskip 0.1cm 
This work is financially supported in part by the Korea Science and Engineering 
Foundation Grant (R01-2000-00015).


\begin{thebibliography}{99}
\bibitem{sample01} R. Hasty et al., Science {\bf 290}, 2117 (2000); R.D. McKeown, Phys. Lett. 
{\bf B219}, 140 (1989).
\bibitem{skyrme61} T.H.R. Skyrme, Proc. Roy. Soc. {\bf A260}, 127 (1961).
\bibitem{witten83npb} G.S. Adkins, C.R. Nappi and E. Witten, Nucl. Phys. {\bf B228}, 552 (1983); 
E. Witten, Nucl. Phys. {\bf B223}, 422 (1983); {\bf B223}, 433 (1983).
\bibitem{brown83prl}  M. Rho, A.S. Goldhaber and G.E. Brown, Phys. Rev. Lett. {\bf 51}, 747 (1983). 
\bibitem{hong98} S.T. Hong, Phys. Lett. {\bf B417}, 211 (1998).
\bibitem{hong02pr} S.T. Hong and Y.J. Park, Phys. Rep. {\bf 358}, 143 (2002).
\bibitem{chodos74}  A. Chodos, R.L. Jaffe, K. Johnson and C.B. Thorn, Phys. Rev. {\bf D10}, 10 (1974). 
\bibitem{gerry791}  G.E. Brown and M. Rho, Phys. Lett. {\bf B82}, 177 (1979).
\bibitem{hong01} D.K. Hong, M. Rho and I. Zahed, Phys. Lett. {\bf B468}, 261 (1999);  
D.K. Hong, S.T. Hong and Y.J. Park, Phys. Lett. {\bf B499}, 125 (2001).
\bibitem{rajagopal00} K. Rajagopal and F. Wilczek, {\it Handbook of QCD}, 
ed. M. Shifman (World Scientific, 2001), and references therein.
\bibitem{thooft74} G. 't Hooft, Nucl. Phys. {\bf B72}, 462 (1974).
\bibitem{dirac64} P.A.M. Dirac, {\it Lectures in Quantum Mechanics} (Yeshiva
University, New York, 1964).
\bibitem{bft} I.A. Batalin and E.S. Fradkin, Phys. Lett. {\bf B180}, 157 (1986);  
I.A. Batalin and I.V. Tyutin, Int. J. Mod. Phys. {\bf A6}, 3255 (1991). 
\bibitem{hong99prd} S.T. Hong and Y.J. Park, Phys. Rev. {\bf D59}, 114026 (1999).
\bibitem{brst} C. Becci, A. Rouet and R. Stora, Ann. Phys. {\bf 98}, 287 (1976); 
I.V. Tyutin, Lebedev Preprint 39 (1975).
\bibitem{hong01prd} S.T. Hong and Y.J. Park, Phys. Rev. {\bf D63}, 054018 (2000).
\bibitem{ccp} M.A. Nowak, M. Rho, I. Zahed, {\it Chiral Nuclear Dynamics} 
(World Scientific, Singapore, 1996), and references therein.
\bibitem{atiyah} M.F. Atiyah, V. Patodi and I. Singer, Math. Proc. Camb. Phil. 
Soc. 77 (1975) 43.
\bibitem{hong93} S.T. Hong and B.Y. Park, Nucl. Phys. {\bf A561}, 525 (1993).
\bibitem{hong93npa} S.T. Hong and G.E. Brown, Nucl. Phys. {\bf A564}, 491 (1993).
\bibitem{coleman}S. Coleman and S.L. Glashow, Phys. Rev. Lett. {\bf 6}, 423 (1961). 
\bibitem{hong97plb} S.T. Hong, B.Y. Park and D.P. Min, Phys. Lett. {\bf B414}, 229 (1997).
\bibitem{kim98} H.C. Kim, M. Praszalowicz, M.V. Polyakov and K. Goeke, Phys. 
                Rev. {\bf D58}, 114027 (1998).
\bibitem{meissner00} Ulf-G. Meissner, Nucl. Phys. {\bf A666}-{\bf A667}, 51 (2000); 
\bibitem{vankolck} C.M. Maekawa and U. van Kolck, Phys. Lett. {\bf B478}, 73 (2000); 
U. van Kolck et al., Phys. Lett. B488 (2000) 167. 
\bibitem{hong01hep} S.T. Hong, {\tt hep-ph/0111470}.
\end{thebibliography}
\end{document}